\documentclass{article}

\PassOptionsToPackage{numbers, compress}{natbib}

\usepackage[final]{neurips_2022}

\usepackage{graphicx}
\usepackage[utf8]{inputenc} 
\usepackage[T1]{fontenc}    
\usepackage{hyperref}       
\usepackage{url}            
\usepackage{booktabs}       
\usepackage{amsfonts}       
\usepackage{nicefrac}       
\usepackage{microtype}      
\usepackage{xcolor}         

\newcommand{\stitle}[1]{\vspace{0.02em}\noindent\textbf{#1}}
\newcommand{\eg}{{\itshape e.g.}, }

\title{Towards Multifaceted Human-Centered AI}

\author{%
  Sajjadur Rahman, Hannah Kim, Dan Zhang, Estevam Hruschka, Eser Kandogan\\
  Megagon Labs\\
\texttt{\{sajjadur,hannah,dan\_z,estevam,eser\}@megagon.ai}
}

\begin{document}

\maketitle

\begin{abstract}
  Human-centered AI workflows involve stakeholders with multiple roles interacting with each other and automated agents to accomplish diverse tasks. In this paper, we call for a holistic view when designing support mechanisms, such as interaction paradigms, interfaces, and systems, for these multifaceted workflows.  
\end{abstract}

\section{Introduction}
Human-centered AI (HCAI) research 
focuses on various aspects: analyzing practitioners' work practices along dimensions such as collaboration~\cite{kandogan2014,zhang2020data}, explainability~\cite{rule2018exploration}, and  trust~\cite{passi2018trust}; conceptualizing specific workflows to inform design guidelines~\cite{muller2019data, passi2017data, pine2015politics,rahman2022ie, wang2019human}; and developing supporting tools and frameworks~\cite{bauerle2022symphony,rahman2020leam,wu2020b2}.
HCAI workflows are multi-faceted, wherein stakeholders with different roles --- \eg product managers, subject matter experts, and data scientists --- perform diverse tasks in various phases using different tools. 
Multiple roles introduce challenges related to collaboration among stakeholders and interaction with tools. The iterative workflows necessitate fine-grained provenance management~\cite{rahman2020leam}. The scope and diversity of domains (\eg NLP, ML, Vision) introduce challenges related to the modeling of domain semantics. To this end, we propose a holistic design that employs plastic interfaces to enable seamless interactions, integrated workflows for ease of task completion, and a graph-based system for managing workflows and  provenance.  

\begin{figure}[!htb] 
\vspace{-5pt}
  \centering
  \includegraphics[width=\linewidth]{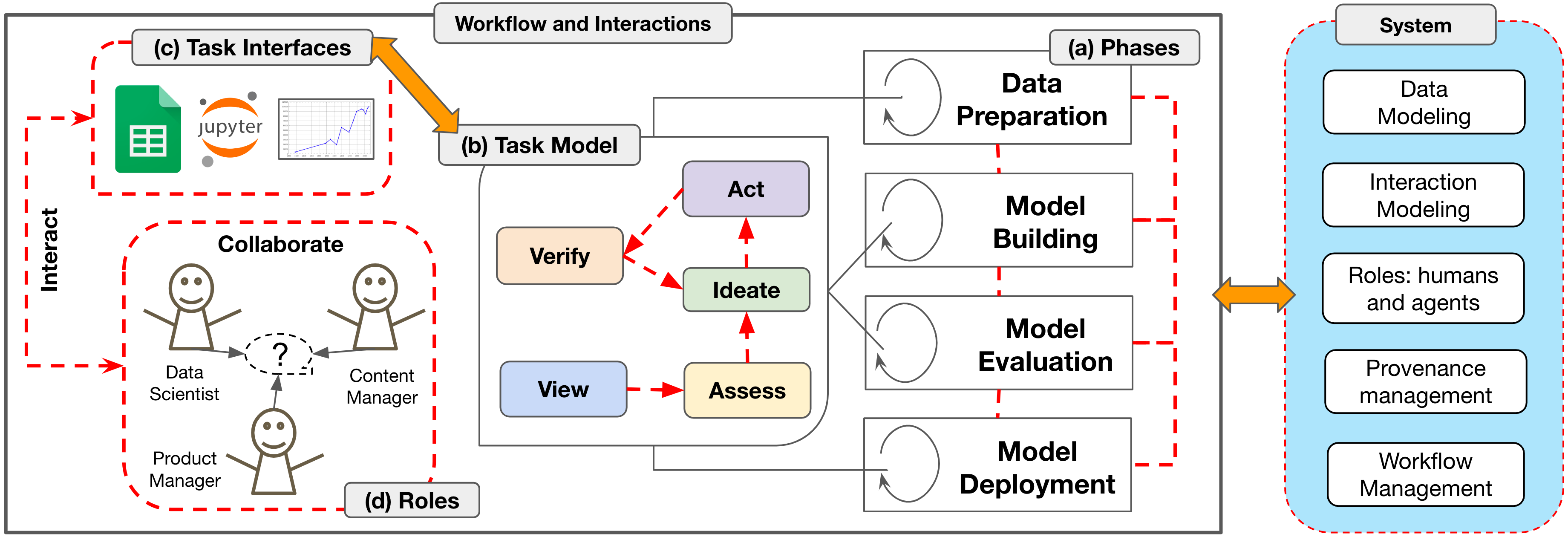}
  \caption{Multifaceted Human-centered AI (M-HCAI) conceptual diagram.}
  \label{fig:mhcai} 
\vspace{-15pt}
\end{figure}

\section{M-HCAI Challenges}
\label{sec:challenges}

Figure~\ref{fig:mhcai} depicts the conceptual diagram of M-HCAI with various workflow components and system requirements.
A workflow is iterative with transitions among \emph{phases} (a). However, a fine-grained characterization, proposed by Rahman et al.~\cite{rahman2022ie}, reveals the iterative nature of a phase with stakeholders performing diverse \emph{tasks} (c). Stakeholders may interact with different tools and interfaces depending on their task requirement, familiarity, and expertise (d)~\cite{rahman2020leam,wongsuphasawat2019goals}. Stakeholders with different roles collaborate in various phases within the workflow both synchronously (in-person and virtual meetings) and asynchronously (\eg Email, Slack, Google Docs)~\cite{rahman2022ie, rahman2020mixtape, zhang2020data}.  

Such complexity of an MHCAI workflow introduces several challenges. First, \emph{iteration}: iterative workflows often force the users to switch multiple tools tediously, often within a single phase~\cite{chattopadhyay2020s,rahman2022ie, rahman2020leam}. Second, \emph{interface rigidity}: existing tools lack the plasticity to accommodate diverse roles and expertise as stakeholders ideate and deliberate~\cite{hong2020human, passi2018trust}. 
Third, \emph{non-standardized collaboration}: as stakeholders collaborate, there may be conflict and misunderstanding~\cite{whiting2019did, whiting2020parallel}. However, there is a lack of standardization around documentation to record these discussions ~\cite{mitchell2019model, rahman2020mixtape, zhang2020data}. Finally, \emph{domain diversity}: when we factor in diverse domains within the HCAI setting such as text~\cite{chiticariu-etal-2010-systemt,rahman2020leam}, image, graphs, and tables~\cite{petersohn13towards}, the lack of formal abstractions and grammar makes it difficult to define and instrument operations in the interfaces systematically.

Another aspect lacking from the HCAI discourse is the management of complicated workflows, which necessitate robust system design. Zhang et al.~\cite{zhang2020data} identified lack of provenance as a contributing factor in obfuscation and loss of knowledge when data science teams share data. Rahman et al.~\cite{rahman2022ie} advocate for instrumenting provenance management mechanisms as built-in features of systems. Kandogan et al.~\cite{kandogan2018context} highlight how existing systems fail to connect and exploit context across stakeholders and agents due to a lack of initiative in modeling data, people, and interactions. 


\section{Multifaceted Human-centered AI: A Holistic View}
We posit that the workflow- and system-level challenges are interdependent. Capturing the provenance of interactions and decisions within the workflow requires provenance-aware interfaces with built-in logging capabilities. For the underlying system to effectively persist and catalog the provenance information, the interactions defined on the interfaces need grounding on sound principles. Such principled design requires (a) building abstractions to capture domain semantics (\eg texts contain words, sentences, POS tags, and opinions), (b) defining operators that represent interactions over those abstractions in an interface (similar to the grammar of interaction graphics~\cite{satyanarayan2016vega}), and (c) explicit modeling of stakeholder roles. Finally, managing these human-centered iterative workflows requires developing frameworks that enable the integration of tools and enhance interface scalability~\cite{bendre2019faster, rahman2020benchmarking}.
In response, we propose the following design considerations for M-HCAI:

\stitle{An integrated workflow supporting multiple paradigms.} Avoiding the context switching overhead during iteration requires integrating different modalities: programming environments and
interactive interfaces (\eg spreadsheets and visualization tools.) Computational notebooks are suitable for such integration where programming can be complemented by introducing interactive widgets~\cite{kery2020mage, wu2020b2}. However, notebooks are not
suitable for non-programmers involved in the HCAI process. Moreover, these tools neither manage the fine-grained provenance of user interactions nor support versioning. Therefore, research efforts for bridging these gaps are crucial for the holistic M-HCAI system design.

\stitle{Plastic interfaces as the basis for interaction.} The multi-role inclusion challenge with computational notebooks can be addressed by developing custom interfaces that may act as boundary objects for the stakeholders. Boundary objects are artifacts that are ``both plastic enough to adapt to local needs and the constraints of the several parties employing them, yet robust enough to maintain a common identity across site'' ~\cite{leigh2010not,star1989structure}. Recent work introduces cross-platform capabilities to transform interactive widgets in notebooks into web dashboards that enable shared understanding among stakeholders within an organizational workflow~\cite{bauerle2022symphony}. However, these widgets are not provenance-aware, and the corresponding interactions are domain semantics agnostic. Therefore, further augmentation is required to employ these interfaces to operate in concert with the underlying system.


\stitle{Systems for supporting M-HCAI.}
Interactions among automated agents and stakeholders of different expertise and roles may occur at any phase. Therefore, the context (where and when a task or interaction occurred), scope (specific data domain and phase), and abstractions of interactions will vary depending on the scenario.
To this end, knowledge graphs~\cite{kg2019def} (KGs) may serve as the core data model for the support systems managing M-HCAI workflows. KGs can explicitly capture diverse stakeholder roles and automated agents, different interaction types, and the corresponding context~\cite{kandogan2018context}. Existing work define grammars to capture operations for text~\cite{chiticariu-etal-2010-systemt, rahman2020leam}, and tabular data analysis~\cite{petersohn13towards}. Capturing domain-specific interaction contexts such as operations and data types would require defining such abstractions for M-HCAI. 

\smallskip \noindent In this position paper, we draw attention to the multi-faceted nature of HCAI while identifying workflow and system-level challenges. We propose a research agenda that requires multi-disciplinary research efforts spanning databases, AI, visualization, and HCI.

\bibliographystyle{abbrvnat}
\bibliography{magneton,position,orchestra}

\end{document}